\documentclass[twocolumn,A4]{article}
\usepackage[dvips]{graphics}
\usepackage{setspace}
\usepackage{color}
\definecolor{gold}{rgb}{0.85,0.66,0}
\definecolor{dblue}{rgb}{0,0,0.5}
\begin{document}
\onecolumn
\begin{center}
{\bf{\Large {\textcolor{gold}{Quantum transport through a molecule 
coupled to a mesoscopic ring: A theoretical study}}}}\\
~\\
{\textcolor{dblue}{Santanu K. Maiti}}$^{1,2,*}$ \\
~\\
{\em $^1$Theoretical Condensed Matter Physics Division,
Saha Institute of Nuclear Physics, \\
1/AF, Bidhannagar, Kolkata-700 064, India \\
$^2$Department of Physics, Narasinha Dutt College,
129 Belilious Road, Howrah-711 101, India} \\
~\\
{\bf Abstract}
\end{center}
Transport through a molecule sandwiched between two metallic electrodes and
coupled to a mesoscopic ring that threads a magnetic flux $\phi$ is studied. 
An analytic approach for the electron transport through the molecular bridge
system is presented. Electronic transport properties are discussed in two
aspects: (a) presence of external magnetic filed and (b) strength of
molecule-to-electrode coupling.
\vskip 1cm
\begin{flushleft}
{\bf PACS No.}: 73.23.-b; 73.63.-b; 81.07.Nb \\
~\\
{\bf Keywords}: Green's function; Quantum transport; Conductance; 
$I$-$V$ characteristic.
\end{flushleft}
\vskip 5.1in
\noindent
{\bf ~$^*$Corresponding Author}: Santanu K. Maiti

Electronic mail:  santanu.maiti@saha.ac.in
\newpage
\twocolumn

\section{{\textsl{Introduction}}}

Much progress in nanofabrication of quantum devices has allowed us to 
study electron transport through molecules in a very controllable way.
Molecular electronics have attracted much more attention since molecules 
constitute promising building blocks for future generation of electronic 
devices. The electron transport through molecules was first studied 
theoretically by Aviram {\em et al.}~\cite{aviram} in $1974$. Later, 
several numerous experiments~\cite{metz,fish,reed1,reed2} have been 
performed through molecules placed between two metallic electrodes with 
few nanometer separation. The operation of such two-terminal devices is 
due to an applied bias. Current passing across the junction is strongly 
nonlinear function of applied bias voltage and its detailed description 
is a very complex problem. The complete knowledge of the conduction 
mechanism in this scale is not well understood even today. Electronic 
transport properties are characterized by several significant factors 
like as the quantization of energy levels, quantum interference of electron 
waves~\cite{mag,lau,baer1,baer2,gold,ern1,walc1,walc2} associated with 
the geometry of the bridging system adopts within the junction and other 
different parameters of the Hamiltonian that are needed to describe the 
complete system. A quantitative understanding of the physical mechanisms 
underlying the operation of nano-scale devices remains a major challenge 
in nanoelectronics research.

The aim of the present article is to reproduce an analytic approach based 
on the tight-binding model to investigate the electronic transport 
properties through a molecule coupled to a mesoscopic ring. There exist 
some {\em ab initio} methods for the calculation of conductance~\cite{yal,
ven,xue,tay,der,dam}, yet it is needed the simple parametric 
approaches~\cite{muj,sam,hjo} for this calculation. The parametric study 
is much more flexible than that of the {\em ab initio} theories since 
the later theories are computationally very expensive and here we do 
attention on the qualitative effects rather than the quantitative ones. 
Not only that, the model calculations by using the tight-binding
formulation also provide a worth insight to the problem. This is why we 
restrict our calculations on the simple analytical formulation of the 
transport problem.

The paper is specifically arranged as follows. In Section $2$, we 
introduce the molecular system under consideration and give a very 
brief description for the calculation of conductance ($g$) and current 
($I$) through the molecular bridge system. Section $3$ presents the 
results of conductance-energy ($g$-$E$) and current-voltage ($I$-$V$) 
characteristics for the bridge system taken into account, and finally, 
we summarize our results in Section $4$.

\section{{\textsl{The molecular model and a brief description onto the 
theoretical formulation}}}

\begin{figure}[ht]
{\centering \resizebox*{7.5cm}{5cm}{\includegraphics{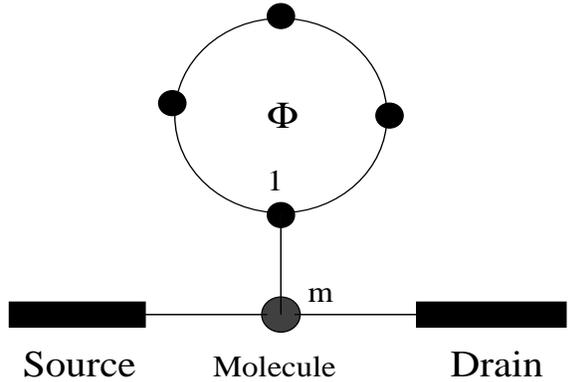}}\par}
\caption{{\textsl{A schematic view of a molecule coupled to a mesoscopic 
ring, threaded by a magnetic flux $\phi$, and the molecule is sandwiched 
between two metallic electrodes.}}}
\label{molecule}
\end{figure}

The system under consideration is a small molecule (with one atomic site) 
coupled to a quantum ring with $N$ atomic sites and attached to two 
one-dimensional metallic electrodes, namely, source and drain, as 
depicted schematically in Fig.~\ref{molecule}.
The bridging system (molecule with ring) between the two electrodes is 
described by a single-band tight-binding Hamiltonian within a 
non-interacting picture, which is written as,
\begin{equation}
H=H_M + H_R + H_{MR}
\end{equation}
\[H_M=\epsilon_{{\mbox{mol}}}\]
\[H_R=\sum_i \epsilon_i c_i^{\dagger} c_i + \sum_{<ij>}t 
\left(c_i^{\dagger}c_j e^{i\theta} + c_j^{\dagger}c_i e^{-i\theta} \right)\]
\begin{equation}
H_{MR}=t_0 \left(c_1^{\dagger}d_m + d_m^{\dagger}c_1 \right)
\label{equ2}
\end{equation}
where $\epsilon_{{\mbox{mol}}}$ is the on-site energy of the molecule and 
$\epsilon_i$'s are the on-site energies of the ring. $c_i^{\dagger}$ ($c_i$) 
is the creation (annihilation) operator of an electron at site $i$ in 
the ring and $d_m^{\dagger}$ ($d_m$) is the creation (annihilation) 
operator of an electron at the molecule (site index $m$). 
$\theta=2 \pi \phi/N$ is the phase factor due to the flux $\phi$ threaded 
by the ring. $t$ is the hopping energy between two nearest-neighbor sites 
in the ring and $t_0$ is the molecule-to-ring tunneling coupling. A
similar kind of tight-binding Hamiltonian, like as in Eq.~(\ref{equ2}), 
is also used to describe the semi-infinite one-dimensional perfect 
electrodes where the Hamiltonian is parametrized by constant on-site
potential $\epsilon^{\prime}$ and nearest-neighbor hopping integral 
$t^{\prime}$.

At low temperatures and bias voltage, the linear conductance of the 
molecular system can be calculated via the one-channel Landauer 
conductance formula through the relation,
\begin{equation}
g=\frac{2 e^2}{h} T
\end{equation}
where $T$ is the transmission probability of an electron from the source to 
drain through the molecular bridge and it can be expressed as~\cite{datta},
\begin{equation}
T(E)={\mbox{Tr}} \left[\left(\Sigma_S^r - \Sigma_S^a\right) 
G^r \left(\Sigma_D^a - \Sigma_D^r\right) G^a\right]
\end{equation}
In this expression, $\Sigma_S$ and $\Sigma_D$ are the self-energies due 
to the coupling of the molecule to the two electrodes. The effective
Green's function $G$ of the bridging system is given by the relation,
\begin{equation}
G=\left[E-H-\Sigma_S-\Sigma_D\right]^{-1}
\end{equation}
where $E$ is the energy of the source electron and $H$ is the Hamiltonian 
of the full system described above. Now all the information regarding the 
molecule-to-electrode coupling are included into the above two 
self-energies and are analyzed through the use of Newns-Anderson 
chemisorption theory~\cite{muj}. The self-energies contain real and 
imaginary parts, where the real parts correspond to the energy shift of 
the energy eigenstates of the molecular system and the imaginary parts 
give the broadening of these energy levels.

The current passing through the bridge is depicted as a single-electron 
scattering process between the two reservoirs of charge carriers. The 
current-voltage relation is evaluated from the following 
expression~\cite{datta},
\begin{equation}
I(V)=\frac{e}{\pi \hbar}\int \limits_{-\infty}^{\infty} 
\left(f_S-f_D\right) T(E) dE
\label{equ5}
\end{equation}
where $f_{S(D)}=f\left(E-\mu_{S(D)}\right)$ gives the Fermi distribution
function with the electrochemical potentials $\mu_{S(D)}=E_F\pm eV/2$.
For the sake of simplicity, here we assume that the entire voltage is 
dropped across the molecule-electrode interfaces and this assumption 
does not greatly affect the qualitative aspects of the $I$-$V$ 
characteristics. 

In this article we perform all the calculations at absolute zero temperature,
but the qualitative behavior of all the results are invariant up to some
finite (low) temperature. The reason for such an assumption is that the
broadening of the energy levels of the molecule including the ring due to 
the coupling to the electrodes is much larger than that of the thermal 
broadening. Throughout the description we set the Fermi energy $E_F$ 
to $0$ and use the unit where $c=e=h=1$.

\section{{\textsl{Results and their interpretation}}}

Here we describe the conductance $g$ as a function of the injecting electron
\begin{figure}[ht]
{\centering\resizebox*{7.5cm}{4.75cm}{\includegraphics{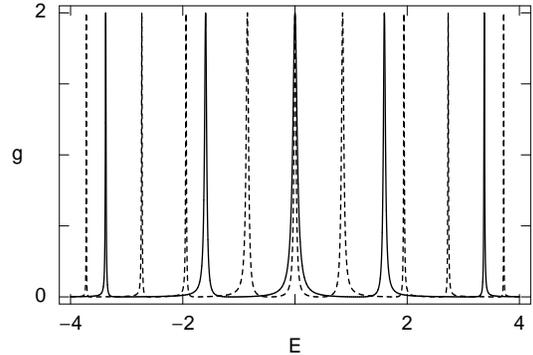}}\par}
\caption{{\textsl{Conductance $g$ as a function of the injecting electron 
energy $E$ in weak-coupling limit, where the solid and dotted curves 
correspond to the results for $\phi=0$ and $0.3$, respectively. The 
size of the ring is fixed at $N=15$.}}}
\label{molcondlow}
\end{figure}
energy $E$ and the variation of the current $I$ with the bias voltage $V$ 
through the molecular bridge system. The transport properties of an 
electron are highly influenced by the molecule-to-electrode coupling 
and here we describe all the essential features of electron transport 
for the two limiting cases of the molecular coupling. One is the
so-called weak-coupling limit defined by the condition $\tau_{\{S,D\}} 
<< t$, and the other one is the strong-coupling limit specified as
$\tau_{\{S,D\}} \sim t$. The parameters $\tau_S$ and $\tau_D$ correspond 
to the couplings of the molecule to the source and drain, respectively. 
The common set of values of these parameters in the two limiting cases 
are as follows: $\tau_S=\tau_D=0.5$, $t=2.5$ (weak-coupling) and 
$\tau_S=\tau_D=2$, $t=2.5$ (strong-coupling). Here we take $t=t_0$ 
(for simplicity). 

The characteristic behavior of the conductance as a function of energy 
in the weak-coupling limit is shown in Fig.~\ref{molcondlow} where, the 
solid and dotted curves correspond to the bridging system with $\phi=0$ and 
\begin{figure}[ht]
{\centering\resizebox*{7.5cm}{4.75cm}{\includegraphics{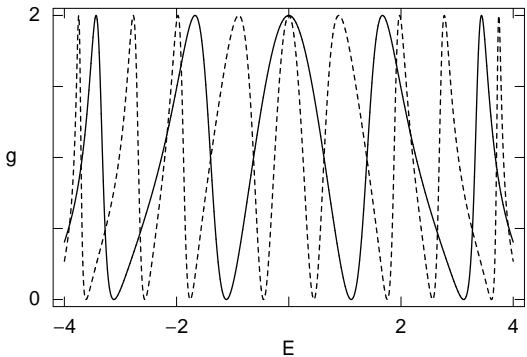}}\par}
\caption{{\textsl{Conductance $g$ as a function of the injecting electron 
energy $E$ in the limit of strong-coupling, where the solid and dotted curves 
correspond to the results for $\phi=0$ and $0.3$, respectively. The size
of the ring is set at $N=15$.}}}
\label{molcondhigh}
\end{figure}
$0.3$, respectively. Conductance vanishes for almost all energy values, 
except at the resonances where it approaches to $2$ i.e., transmission 
probability $T$ goes to unity since $g=2T$ (from Landauer formula with 
$e=h=1$). The resonant peaks in the conductance spectrum coincide with 
eigenenergies of the molecular system and thus the spectrum manifests 
itself the energy levels of the molecular bridge system. In the presence 
of magnetic flux $\phi$, more resonant peaks appear in the conductance 
spectrum (dotted curve) which reveals that more energy levels appear in 
the system. The reason is that for the non-zero value of $\phi$, all 
the degeneracies of the molecular energy levels are lifted and eventually 
the system gets more resonating states. Thus by introducing $\phi$, the
transmission through the bridge system can be controlled. 

In the strong-coupling limit, these resonant peaks get substantial widths 
as observed in Fig.~\ref{molcondhigh}, where the solid and dotted curves 
correspond to the same meaning as earlier. The increment of the resonant 
widths is due to the broadening of the molecular energy levels, where
the contribution comes from the imaginary parts of the two self-energies 
in the strong molecular coupling to the electrodes. 

The scenario of electron transfer through the molecular bridge system is 
much more clearly observed by studying the current-voltage characteristics.
The current is computed by the integration procedure of the transmission 
\begin{figure}[ht]
{\centering\resizebox*{7.5cm}{4.75cm}{\includegraphics{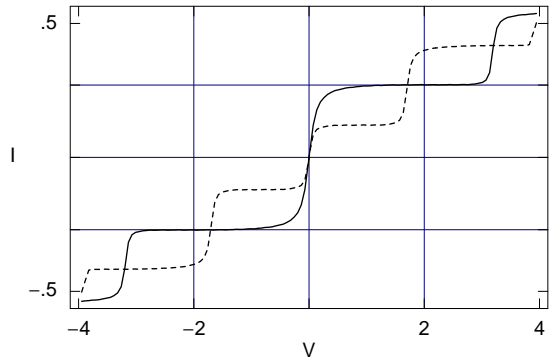}}\par}
\caption{{\textsl{Current $I$ as a function of the applied bias voltage 
$V$ in the weak molecule-to-electrode coupling limit, where the solid 
and dotted lines correspond to the results for $\phi=0$ and $0.3$, 
respectively. The parameter $N$ is fixed at $15$.}}}
\label{molcurrlow}
\end{figure}
function $T$ which shows the same variation, differ only in magnitude by 
the factor $2$, like as the conductance spectrum (Figs.~\ref{molcondlow} 
and \ref{molcondhigh}). The current-voltage characteristic in the 
weak-coupling limit for the molecular system is given in 
Fig.~\ref{molcurrlow}, where the solid curve corresponds to the current 
in the absence of any magnetic flux, while the dotted curve denotes the 
same for the flux $\phi=0.3$. The current shows staircase-like behavior 
with sharp steps, which is associated with the discrete nature of the 
molecular resonances (Fig.~\ref{molcondlow}). It is clearly observed 
\begin{figure}[ht]
{\centering\resizebox*{7.5cm}{4.75cm}{\includegraphics{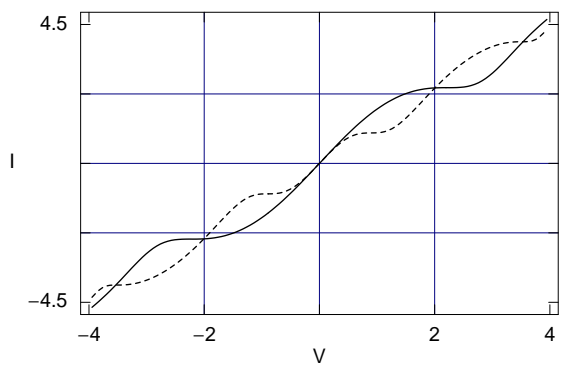}}\par}
\caption{{\textsl{Current $I$ as a function of the applied bias voltage 
$V$ in the strong molecule-to-electrode coupling limit, where the solid 
and dotted lines correspond to the results for $\phi=0$ and $0.3$, 
respectively. The parameter $N$ is taken as $15$.}}}
\label{molcurrhigh}
\end{figure}
from this figure that, in the presence of magnetic flux, current shows 
more steps (dotted curves) compared to the case when $\phi$ becomes zero.
It reveals that more resonant peaks appear in the conductance spectrum 
for the non-zero value of $\phi$. The shape and width of the current steps 
depend on the width of the molecular resonances since the hight of a step 
in $I$-$V$ curve is directly proportional to the area of the corresponding 
peak in the conductance spectrum. The current varies continuously with the 
applied bias voltage and achieves much higher values in the strong-coupling 
limit, as shown in Fig.~\ref{molcurrhigh}, where the solid and dotted 
curves correspond to the same meaning as earlier.

\section{Concluding remarks}

To summarize, we have introduced a parametric approach based on the 
tight-binding model to investigate the electronic transport properties at 
absolute zero temperature through a single molecule coupled to a mesoscopic 
ring that threads a magnetic flux $\phi$. This method is much more 
flexible compared to other existing theories and can be used to study 
electron transport properties through any complicated bridge system. 
Electronic conduction through the molecular systems is strongly influenced
by the flux $\phi$ in the ring and the molecule-to-electrode coupling 
strength. These results predict that designing a whole system that 
includes not only molecules but also molecule-to-electrode coupling is 
highly important in fabricating molecular electronic devices.

\end{document}